\documentclass{emulateapj}
\usepackage{graphicx}
\begin{document}

\def\ba{\begin{eqnarray}}
\def\ea{\end{eqnarray}}
\def\etal{et al.\ \rm}

\title{Cooling of young stars growing by disk accretion.}

\author{Roman R. Rafikov\altaffilmark{1,2}}
\altaffiltext{1}{CITA, McLennan Physics Labs, 60 St. George St., 
University of Toronto, Toronto ON M5S 3H8 Canada; rrr@cita.utoronto.ca}
\altaffiltext{2}{Canada Research Chair}


\begin{abstract}
In the initial formation stages young stars must 
acquire a significant fraction of their mass by accretion from a 
circumstellar disk that forms in the center of a collapsing protostellar 
cloud. Throughout this period mass accretion rates through the disk 
can reach $10^{-6}-10^{-5}$ M$_\odot$ yr$^{-1}$ leading to 
substantial energy release in the vicinity of stellar surface.
We study the impact of irradiation of the stellar surface produced 
by the hot inner disk on properties of accreting fully 
convective low-mass stars, and also look at objects such as young 
brown dwarfs and giant planets. At high accretion rates irradiation 
raises the surface temperature of the equatorial region 
above the photospheric temperature $T_0$ that a star would 
have in the absence of accretion. The high-latitude (polar) 
parts of the stellar surface, where disk irradiation is weak, 
preserve their temperature at the level of $T_0$. In strongly 
irradiated regions an almost isothermal outer radiative zone 
forms on top of the fully convective interior, leading
to the suppression of the local internal cooling flux derived from 
stellar contraction (similar suppression occurs in 
irradiated ``hot Jupiters''). Properties of this radiative 
zone likely determine the amount of thermal energy that gets advected 
into the convective interior of the star. 
Total intrinsic luminosity integrated over the whole stellar 
surface is reduced compared to the non-accreting case, by 
up to a factor of several in some systems (young brown dwarfs, 
stars in quasar disks, forming giants planets),
potentially leading to 
the retardation of stellar contraction.
Stars and brown dwarfs irradiated by their disks tend to lose energy 
predominantly through their cool polar regions while young giant 
planets accreting through the disk cool through their whole surface.
\end{abstract}


\section{Introduction.}
\label{sect:intro}


Our understanding of advanced stages of star formation (T Tauri
and later phases) has been significantly
improved with the advent of infrared, 
submillimeter, and high-resolution optical observatories
such as HST and Spitzer. At the same time a great deal of 
uncertainty still remains regarding the earliest, so-called
Class 0 and Class I, stages of the star formation process. In the 
conventional nomenclature, Class 0 stars are the protostellar
cloud cores in the very beginning of their collapse, while
Class I are the protostars embedded within an envelope of circumstellar 
material that is infalling, accumulating in the centrifugally
supported disk, and being accreted by the protostars. 

At present, our knowledge exhibits a significant gap when 
it comes to describing the actual buildup of the stellar 
mass, from $M_\star=0~M_\odot$
in the Class 0 phase to $M_\star\sim 1~M_\odot$ in the end of
Class I phase. From the observational point of view the major 
reasons for this 
are (1) the heavy obscuration provided by the increased 
densities in the central part of the infalling  protostellar
core and the molecular cloud as a whole and (2) the 
difficulty in deriving the spectra of the
central objects, namely distinguishing between the 
intrinsic protostellar and accretion luminosities.
At the same time, our ignorance concerns not only the 
history of {\it mass} 
accumulation by the protostars. Thermodynamical state of the
accumulated gas is also an important ingredient of the 
picture. Stars that form out of material with high entropy, 
in particular that processed through 
the accretion shock, tend to have large sizes, while objects 
formed out of the lower entropy gas should be more compact.
At the moment the uncertainty in the initial 
thermodynamical state of protostellar objects 
precludes us from 
getting a good handle on the evolutionary tracks of the fully
assembled (in terms of mass) protostars in the first $1-10$
Myrs after their formation. 
Beyond 
about 10 Myrs, when the initial conditions become largely 
forgotten, the evolution tracks calculated under
different assumptions about the initial conditions typically
converge (Baraffe \etal 2002). However, prior to this stage 
there are significant discrepancies between the results of
different groups, and the uncertainty in the initial 
conditions for such calculations is the prime suspect for 
the difference. 

It is generally accepted that the conservation of the angular 
momentum in the collapsing protostellar cloud results in 
accumulation of the collapsed gas in a rotationally-supported 
disk in the cloud center. Only a small fraction 
of the cloud mass has low enough angular momentum to 
collapse directly into the protostellar core. The majority of  
stellar mass is most likely accumulated by accretion from the 
disk. According to observations, Class I stars acquire most 
of their mass on timescale of several $10^5$ yrs
which implies that disks around these objects must exhibit time averaged 
mass accretion rates of $\dot M \sim 10^{-6}-10^{-5}$ 
M$_\odot$ yr$^{-1}$. Accretion luminosity released in processing 
such large mass flux through the disk exceeds the intrinsic
luminosity of the protostar. This immediately raises 
an issue of the possible non-trivial radiative coupling between 
the protostar and its circumstellar disk. 

\begin{figure}
\plotone{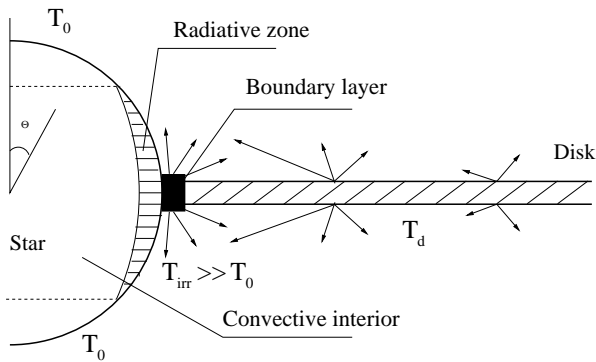}
\caption{
Schematic representation of stellar illumination  by the disk
(slant hashed). Filled region at the point where disk joins the star
marks the boundary layer where intense energy dissipation 
takes place. Part of radiation emitted by the disk (arrows) gets 
intercepted by the star which heating it to temperature $T_{irr}$, 
higher than the temperature $T_0$ that a star would have in 
the absence of irradiation. Photospheric temperature is
preserved at the level of $T_0$ only in the polar regions
of the star (marked with dashed lines) where disk illumination
is weak. An external radiative zone 
(horizontally hashed) forms in the strongly irradiated parts 
of the stellar surface.
\label{fig:scheme}}
\end{figure}

Effects of disk accretion on structure of young stars 
have been investigated by Mercer-Smith \etal (1984), 
Palla \& Stahler (1992), Siess \& Forestini (1996),
Hartmann \etal (1997), Siess \etal (1997, 1999). 
Some of these authors studies how the heat advected 
into the star with the freshly accreted material affects 
protostellar properties. However, none of these investigations 
looked at the effect of heat deposited at the stellar 
{\it surface} by radiation originating in the inner parts 
of the circumstellar disk, where most of the accretion energy 
is released (see Figure \ref{fig:scheme} for a schematic 
representation). Given that accretion luminosity may easily 
exceed the intrinsic stellar luminosity (luminosity derived from 
gravitational contraction, cooling and, possibly, deuterium 
burning in stellar interior), omission of this effect may not 
be justified in many cases.

In this paper we investigate stellar irradiation by the 
circumstellar disk and address the importance of this effect 
in determining the intrinsic luminosity of young stars. We 
calculate the spatial distribution of the disk flux on
the stellar surface and determine when irradiation is
important in \S \ref{sect:T_dist}. The effect of irradiation
on stellar cooling is investigated locally in \S 
\ref{sect:stellar_cool} and globally in \S 
\ref{sect:total_cooling}. Finally, in \S \ref{sect:disc} 
we discuss the applications of this study to some real 
objects and its possible limitations.


\section{Temperature distribution due to disk irradiation.}
\label{sect:T_dist}


We start by calculating the distribution on the stellar surface 
of the radiative flux $F_{irr}$ produced by the disk.
We consider an axisymmetric geometrically thin disk accreting 
onto a star with radius $R_\star$ and mass $M_\star$.  
Flux $F_{irr}$ intercepted by the star is a function of $\theta$ 
-- the angle between the normal to the stellar surface and the 
normal to the disk (coincident with the polar axis of the star, 
assuming that disk lies in the stellar equatorial plane). 
Polar regions of the star are exposed to the radiation 
of only the distant, cool parts of the disk, while the equatorial 
regions have a direct view to the innermost parts of the disk 
where most of the energy is dissipated.
One can easily show that a disk extending all the way to the 
stellar surface gives rise to irradiation flux $F_{irr}(\theta)$ 
given by (Adams \& Shu 1986; Popham 1997)
\ba
&& F_{irr}(\theta)=2\frac{R_\star\cos\theta}{\pi}
\int\limits_{R_{in}}^\infty F_d(R)R dR\nonumber\\
&& \times\int\limits_0^{\phi_c} d\phi\frac{R\sin\theta\cos\phi-R_\star}
{(R^2+R_\star^2-2R_\star R\sin\theta\cos\phi)^2}
\label{eq:irr_flux}
\ea
where $R$ is the cylindrical radius, 
$\cos\phi_c=R_\star/(r\sin\theta)$, $R_{in}=R_\star/\cos\theta$, 
and $F_d(R)$ is the energy radiated by the unit surface area 
of the disk per unit of time. In Appendix A we demonstrate 
that this expression can be reduced to a one-dimensional 
integral which is easier to analyze than equation 
(\ref{eq:irr_flux}). 

To find the explicit dependence of $F_{irr}$ on $\theta$ one 
needs to know $F_d(R)$ which is determined by the viscous 
dissipation in the disk. Studies of steady-state thin accretion disks 
have generally found that
\ba
F_d(R)=\frac{3}{8\pi}\frac{G M_\star \dot M}{R^3}f(R)
\label{eq:vis_dissip}
\ea
where $\dot M$ is a mass accretion rate and the function $f(R)$,
embodying the details of the disk emissivity near the stellar 
surface, behaves as $f\to 1$ when $R\gg R_\star$. With $F_d$ 
given by (\ref{eq:vis_dissip}) one finds
\ba
F_{irr}(\theta)=\frac{G M_\star \dot M}{R_\star^3}g(\theta),
\label{eq:irr_flux_mod}
\ea
where the dimensionless function $g(\theta)$ is given by equation 
(\ref{eq:g}).

A standard disk with zero torque at the stellar surface 
(situation appropriate for accretion onto  
black holes) has (Shakura \& Sunyaev 1973) 
$f(R)=1-(R_\star/R)^{1/2}$. The total viscous dissipation in 
such a disk is $\dot E_d = (1/2)GM_\star \dot M/R_\star$ 
and the gas at the inner edge of the disk rotates at the local Keplerian 
velocity. This is inappropriate in our case since the gas speed 
has to match the velocity of the stellar surface at $R=R_\star$
(for simplicity assumed to be zero in our case).
As a result a boundary layer must form near the stellar surface in 
which the azimuthal velocity of the gas is lowered by the viscous 
torque from the local Keplerian value to the stellar rotation speed. 
Viscous dissipation dramatically 
increases gas temperature in this layer creating an additional 
source of radiative flux very close to the stellar surface. 
Irradiation by the boundary layer emission boosts up the
stellar surface temperature in a narrow belt at the equator (with
the thickness in $\theta$-direction comparable to the thickness
of the boundary layer) above that expected from the irradiation 
by the more distant parts of the disk, outside of the boundary layer.
Thus, the existence of the boundary layer significantly modifies 
disk structure and emissivity near the stellar surface
(Popham \etal 1993; Popham \& Narayan 1995) complicating 
the calculation of $f(R)$.

\begin{figure}
\plotone{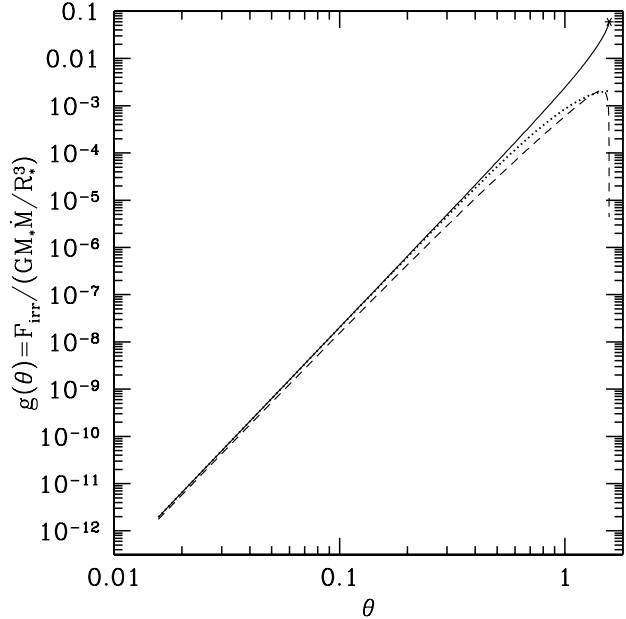}
\caption{
Irradiation flux absorbed by the stellar surface in units of 
$GM_\star \dot M/R_\star^3$ as a function of $\theta$. Solid
curve corresponds to 
$F_d(R)\propto R^{-3}$, dashed curve corresponds to  
$F_d(R)\propto R^{-3}[1-(R_\star/R)^{1/2}]$, while the dotted 
curve represents asymptotic behavior (\ref{eq:as}).
\label{fig:irr_flux}}
\end{figure}

Fortunately, it will be shown later in \S 
\ref{sect:total_cooling} that cooling of irradiated stars
depends only weakly on the behavior of $f(R)$ at $R\sim R_\star$
and is thus relatively insensitive to the structure of the 
boundary layer. For the mass 
accretion rates considered in this work ($10^{-6}-10^{-5}$ 
M$_\odot$ yr$^{-1}$) the geometric thickness of the boundary 
layer is rather small, $\lesssim 0.15$ R$_\star$ (Popham \etal 
1993), so that the fraction of the stellar surface covered by 
the boundary 
layer and affected by the energy dissipation in it is rather small. 
For this reason we will further assume for simplicity 
that\footnote{Such assumption 
results in $\dot E_d = (3/2)GM_\star \dot M/R_\star$, larger 
than what can be provided by the change of the potential energy of 
disk material, but this inconsistency is not going to strongly 
affect our results.} $f(R)\approx 1$. In this case 
$F_d$ keeps increasing all the way to the stellar surface [unlike 
the zero inner torque case in which $F_d(R_\star)\to 0$] 
thus roughly mimicking the contribution of the boundary layer to
the disk flux. We plot the behavior of function $g(\theta)$ in Figure 
\ref{fig:irr_flux} for both $f(R)=1$ and $f(R)=1-(R_\star/R)^{1/2}$.

Irrespective of the complications related to the existence of 
the boundary layer one can derive useful results for
$F_{irr}(\theta)$ in two asymptotic regimes. In particular, 
using equation (\ref{eq:1D}) one finds that 
as $\theta\to\pi/2$ 
\ba
F_{irr}\to\frac{F_d(R_\star)}{2},~~~~~g\approx\frac{3}{16\pi},
\label{eq:T_pi_2}
\ea
a result that is easy to 
understand since any point at the stellar equator receives 
disk radiation with uniform temperature corresponding to local
disk flux $F_d(R_\star)$
from $\pi$ steradian and reemits it into $2\pi$ steradian.
In the other limit
of $\theta\to 0$ one finds from equation (\ref{eq:1D}) that
\ba
g\approx I_1\sin^5\theta,
\label{eq:as}
\ea
where constant $I_1$ is given by equation (\ref{eq:I}). One can
see from Figure \ref{fig:irr_flux} that approximation (\ref{eq:as}) 
works 
quite well (better than $22\%$ accuracy) for $\theta\lesssim 0.5$.
This asymptotic behavior is insensitive to the details of the 
disk emissivity at $R\sim R_\star$ since  polar regions 
are irradiated only by parts of the disk at $R\gg R_\star$ where 
$f(R)\approx 1$. 

Let us denote $T_0$ and $L_0=4\pi R_\star^2\sigma T_0^4$ 
the temperature and luminosity which a star with mass 
$M_\star$ and radius $R_\star$ would have in the absence of 
irradiation ($\sigma$ is a Stephan-Boltzmann constant). 
To characterize the importance of irradiation we introduce 
{\it irradiation parameter} $\Lambda$:
\ba
\Lambda\equiv\frac{GM_\star\dot M}{\sigma T_0^4 R_\star^3}\approx
1.6~M_1\dot M_{-9}R_{11}^{-3}T_{3.5}^{-4},
\label{eq:Lambda}
\ea
where $T_n\equiv T_0/10^n$ K, $R_n\equiv R_\star/10^n$ cm, 
$M_1\equiv M_\star/M_\odot$, and 
$\dot M_n\equiv \dot M/(10^n~M_\odot~ {\rm yr}^{-1})$.
By construction, $\Lambda$ is roughly the ratio of accretion 
luminosity and the stellar luminosity $L_0$ in the absence 
of irradiation.

When irradiation is 
allowed for the photospheric temperature of the star $T_{ph}$ 
is a function of $\theta$ since 
energy balance in steady state requires 
\ba
\sigma T_{ph}^4(\theta) = \sigma T_{irr}^4(\theta) + F_{in}
\label{eq:balance}
\ea
at each point on the stellar surface, 
where $F_{in}$ is 
the intrinsic energy flux coming from the stellar interior 
(derived from cooling of the stellar interior, 
gravitational contraction, and D burning) and
\ba
&& T_{irr}(\theta)\equiv \left[\frac{F_{irr}(\theta)}
{\sigma}\right]^{1/4}=T_0\left(\Lambda g\right)^{1/4}\\ 
&& \approx 3.5\times 10^4  ~\mbox{K}
\left(M_1\dot M_{-5}R_{11}^{-3}\right)^{1/4}g^{1/4}.
\nonumber
\label{eq:T_irr}
\ea
Equation (\ref{eq:balance}) assumes that all radiation intercepted 
by the star gets fully absorbed by its surface and reflection is 
negligible. Our discussion can be easily extended
for the case of non-zero stellar albedo.

In the absence of irradiation $F_{in}=\sigma T_0^4$.  
With irradiation the local flux emitted by the photosphere
$\sigma T_{ph}^4$ exceeds $\sigma T_0^4$, but the 
intrinsic stellar flux $F_{in}$ derived from the gravitational 
contraction and cooling of the stellar interior
actually becomes smaller than $\sigma T_0^4$ as we 
demonstrate in \S \ref{sect:1D}.

We define the regime of {\it weak} irradiation as that 
corresponding to low $\dot M$ such that 
\ba
T_{irr}(\theta)\lesssim T_0
\label{eq:weak_condition}
\ea
for any $\theta$ (i.e. $\Lambda g\ll 1$). As the irradiation 
is more intense near the stellar equator, weak irradiation
at {\it any} point on the stellar surface requires 
$T_{irr}(\pi/2)\lesssim T_0$ (or $\Lambda\lesssim 1$), 
or accretion rates lower than 
\ba
&& \dot M_c\approx \frac{16\pi}{3}\frac{R_\star^3\sigma T_0^4}
{G M_\star}\nonumber\\
&& \approx 10^{-8}T_{3.5}^4 R_{11}^3 M_1^{-1}~
{\rm M}_\odot ~{\rm yr}^{-1}.
\label{eq:crit_M_dot}
\ea
Energy dissipation in the equatorial boundary layer (which we 
do not account for here) can heat equatorial region  
above $T_0$ even at $\dot M\lesssim  \dot M_c$, but this heating 
does not spread very far from the equator and 
does little to affect the large scale stellar structure.

A regime of {\it strong} irradiation is defined as that 
corresponding to $\dot M\gtrsim \dot M_c$ ($\Lambda\gg 1$) 
so that at least some parts of the stellar surface have 
\ba
T_{irr}(\theta)\gtrsim T_0
\label{eq:strong_condition}
\ea
(or $\Lambda g \gtrsim 1$).
Initially this condition is satisfied only near the stellar 
equator where an irradiated belt with 
$T_{irr}(\theta)\gtrsim T_0$  forms. As $\dot M$ increases 
this belt expands in $\theta$-direction, although rather 
slowly since $F_{irr}$ is a rapidly decreasing function of 
$\theta$, see Figure \ref{fig:irr_flux}. 
As will be shown in \S \ref{sect:1D}, in irradiated regions 
the intrinsic energy flux $F_{in}$ coming from the stellar 
interior is suppressed compared 
to $\sigma T_0^4$ so that the effective temperature of the 
irradiated part of the star can be well approximated by 
\ba
T_{ph}\approx T_{irr}(\theta).
\label{eq:irr_temp}
\ea
Transition between the low-latitude irradiated region 
and the high-latitude part of the stellar surface where 
$T_{ph}\approx T_0$ takes place at $\theta_{irr}$ given 
by (see eq. [\ref{eq:as}])
\ba
&&\sin\theta_{irr}\approx \left(\frac{R_\star^3\sigma T_0^4}
{I_1 G M_\star \dot M}\right)^{1/5}=(I_1\Lambda)^{-1/5}\nonumber\\
&&\approx 0.5~
T_{3.5}^{4/5} R_{11}^{3/5} M_1^{-1/5}\dot M_{-5}^{-1/5}.
\label{eq:theta_c}
\ea
According to this formula, at $\dot M=10^{-5}$ M$_\odot$ 
yr$^{-1}$ polar regions having temperature $T_0\approx 3000$ K  
occupy about $15\%$ of the stellar surface. The rest of 
the surface has $T_{ph}$ 
significantly modified by intense radiation coming 
from the disk. At this $\dot M$ equatorial temperature 
reaches $T_{ph}(\pi/2)\approx 1.8\times 
10^4$ K, much higher than $T_0\sim 3000$ K corresponding 
to the typical Hayashi track of a young star.


\section{Cooling of irradiated stellar surface.}
\label{sect:stellar_cool}


Young stars, brown dwarfs and giant planets are fully 
convective objects. It is well known (Kippenhahn \& Weigert 1994) 
that many characteristics of fully convective objects such as 
their luminosity and effective temperature are determined 
mainly by the properties (opacity behavior, ratio of specific 
heats of the gas) of their outermost, near-photospheric layers 
and are rather insensitive to the processes 
occurring in the convective interior.
Given that irradiation changes the boundary conditions
on the surface of accreting fully convective object we may
also expect that it should affect the luminosity of such an 
object (Arras \& Bildsten 2006).

Intense heating some parts of the 
stellar surface suppresses convection in the subsurface layers
and gives rise to a convectively stable radiative zone sandwiched 
between the photosphere and convective interior, see Figure 
\ref{fig:scheme} for illustration. Appearance 
of this zone is analogous to the formation of a roughly isothermal 
radiative layer in the outer parts of the close-in extrasolar giant planets 
caused by the intense radiation of their parent stars 
(Guillot \etal 1996; Burrows 2000). It is the structure of 
this zone that we want to investigate in order  
to assess an impact of irradiation on stellar cooling.
Here we assume that the radiative zone is
\begin{enumerate}
\item optically thick, as measured from its bottom 
(convective-radiative boundary) to the photosphere, and
\item geometrically thin compared to $R_\star$.
\end{enumerate}
Validity of these assumptions is verified in \S \ref{sect:1D}.

In the absence of internal energy sources radiation 
transport in the optically thick radiative layer is 
governed by 
\ba
\nabla\cdot {\bf F}=0,~~~~{\bf F}=
-\frac{16}{3}\frac{\sigma T^3}{\kappa\rho}\nabla T,
\label{eq:rad_tran}
\ea
where $F$ is the radiative flux density, $\kappa$ is opacity 
and $\rho$ is the gas density. Equation of hydrostatic equilibrium 
reads $\nabla P = -\rho {\bf g}$, where $P$ is the gas pressure and
${\bf g}$ is the local gravitational acceleration. These two equations
describe the radiative zone structure subject to the boundary condition
\ba
&& T\Big|_{\tau=2/3}=T_{ph}(\theta),
\label{eq:Tboundary}
\ea
where $\tau$ is the optical depth. The radiative boundary condition 
(\ref{eq:Tboundary}) coupled with (\ref{eq:balance}) is appropriate 
here because most of the stellar surface is not obscured by 
the accreting gas and is free to radiate energy into space.


\subsection{1D approximation for the structure of the radiative layer.}
\label{sect:1D}

In general one must solve equation (\ref{eq:rad_tran}) together with 
the equation of hydrostatic equilibrium in two dimensions -- $r$ and 
$\theta$. However, under the circumstances clarified in \S 
\ref{sect:1D_validity} the $r$-component of the radiative
flux $F_r$ is much larger than its $\theta$-component $F_\theta$, 
so that the latitudinal transport of energy can be neglected.
This leaves $r$ as the only independent variable in equation 
(\ref{eq:rad_tran}) effectively making it one-dimensional.
A dependence on $\theta$ then appears only through the external 
boundary condition, namely $T_{ph}(\theta)$. This is the limit that 
we will focus on in this work. 

Because of the thinness of the radiative zone, $r$ varies 
only weakly through the radiative zone so that one can neglect 
the divergence of the radial component of $F_r$, thereby 
reducing equation (\ref{eq:rad_tran}) to simply 
$\partial F_r/\partial r =0$.
With these simplifications equation (\ref{eq:rad_tran}) can be
integrated once to find
\ba
F_{in}=-\frac{16}{3}\frac{\sigma T^3}
{\kappa\rho}\frac{\partial T}{\partial r},
\label{eq:flux}
\ea
where the integration constant on the left-hand side is 
independent of $r$ and as such has to coincide with the 
intrinsic flux $F_{in}$ coming from 
the convective interior of the star. Determination of 
$F_{in}$ is the goal of our calculation.

Radial pressure gradients in the radiative zone
are much larger than the latitudinal gradients so that the 
equation of hydrostatic equilibrium can be written as
\ba
\frac{\partial P}{\partial r} = -\rho g,
\label{eq:hydro_eq}
\ea
where $g=|{\bf g}|\approx GM_\star/R_\star^2$ is the gravitational 
acceleration which is roughly constant within the thin radiative 
layer (stellar rotation is neglected throughout this work).

Subsequent consideration is very similar to the calculation of 
the atmospheric structure for the protoplanetary core 
immersed in a protostellar nebula, which can be found in Rafikov 
(2006). We assume that $\kappa$ depends on
gas pressure and temperature as
\ba
\kappa=\tilde\kappa P^\alpha T^\beta,
\label{eq:kappa}
\ea
where $\alpha>0$ and $\beta$ are constants. 
Equation (\ref{eq:kappa}) is a reasonable approximation to the
opacity behavior in some density and temperature intervals 
typical for young stars. In particular, 
at $2500$ K $<T\lesssim 5000$ K opacity is mainly 
due to H$^-$ absorption with electrons supplied by
elements heavier than H with low ionization potentials. 
Bell \& Lin (1994) have demonstrated that in this regime  
$\kappa$ can be well fit by  
\ba
\kappa\approx 6\times 10^{-14}P^{2/3}T^{7/3}.
\label{eq:low_T}
\ea
At $T\gtrsim 5000$ K electrons from partial hydrogen ionization 
enhance H$^-$ opacity and hydrogenic absorption dominates. In this regime 
(Bell \& Lin 1994)
\ba
\kappa\approx 2.4\times 10^{-39}P^{1/3}T^{29/3}.
\label{eq:high_T}
\ea

Here we also assume that the equation of state (EOS)
of gas in the whole star, including the external radiative zone,
can be characterized by a single ratio of specific heats $\gamma$. 
In other words, we assume that under adiabatic conditions gas behaves as 
$P=K\rho^\gamma$, where $K$ is a constant set by the entropy of the gas 
and $\gamma$ is fixed throughout the star. We adopt $\gamma=5/3$, which 
should work fine in fully ionized, convective interiors of young low-mass 
stars, although this approximation is not very accurate at the transition 
between the outer radiative zone and the convective interior since 
gas is only partly atomic there. Continuing 
dissociation and ionization cause variations of $\gamma$ in this region 
which may be quite important, see \S \ref{sect:complications}. 
Nevertheless, to get a qualitative picture of the effect of irradiation 
on stellar cooling and for making rough numerical estimates this 
constant-$\gamma$ approximation should be sufficient.

With $\kappa$ in the form (\ref{eq:kappa}) 
equation (\ref{eq:flux}) can be integrated using 
(\ref{eq:hydro_eq}), (\ref{eq:kappa}) and the 
ideal gas law:
\ba
\left(\frac{P}{P_{ph}}\right)^{1+\alpha}-1 = \frac{\nabla_0}{\nabla_{ph}}
\left[\left(\frac{T}{T_{ph}}\right)^{4-\beta}-1\right],
\label{eq:PofT}
\ea
where 
\ba
&& \nabla_0=\frac{1+\alpha}{4-\beta},
\label{eq:nabla_0}\\
&& \nabla_{ph}= \frac{3}{16}\frac{F_{in}\kappa_{ph}P_{ph}}
{g\sigma T_{ph}^4},
\label{eq:nabla_ph}
\ea
and $P_{ph}$ and $\kappa_{ph}=\tilde \kappa P_{ph}^\alpha
T_{ph}^\beta$ are the values of pressure and opacity at the 
photosphere. 

Solution (\ref{eq:PofT}) allows us to calculate temperature gradient 
\ba
\nabla(T)\equiv\frac{\partial \ln T}{\partial \ln P}=\nabla_0
\left[1-\left(\frac{T_{ph}}{T}\right)^{4-\beta}
\left(1-\frac{\nabla_{ph}}{\nabla_0}\right)\right],
\label{eq:nabla}
\ea
which determines whether gas is stable against convection.
Note that at the photosphere $\nabla(T_{ph})=\nabla_{ph}$.
Everywhere inside the radiative zone 
\ba
\nabla<\nabla_{ad}\equiv 
(\gamma-1)/\gamma,
\label{eq:Schwar}
\ea
where $\nabla_{ad}$ is the adiabatic temperature gradient. 
In convective regions $\nabla>\nabla_{ad}$. For our adopted 
$\gamma=5/3$ one finds $\nabla_{ad}=2/5$. 

We also assume that at some depth an object under consideration
does become convective and determine what is necessary for this 
transition to occur. If $\beta<4$ then equations (\ref{eq:nabla}) 
and (\ref{eq:Schwar}) demonstrate that convection sets
in only provided that
\ba
\nabla_0>\nabla_{ad}. 
\label{eq:less4}
\ea
Situation described by equation (\ref{eq:less4}) is realized e.g. 
for opacity given by equation (\ref{eq:low_T}), when $\nabla_0=1$ 
exceeds $\nabla_{ad}=2/5$, implying that radiative energy 
transport does indeed change to convective at some depth, 
as we have assumed. 

On the other hand, when opacity is 
characterized by $\beta>4$ equation (\ref{eq:nabla}) guarantees 
that transition to convection occurs at some depth, irrespective 
of the exact value of either $\beta$ or $\nabla_0$. This situation 
is appropriate for $\kappa$ given by equation (\ref{eq:high_T}) 
since in that case $\beta\approx 10$.

Despite this difference, in both cases the temperature $T_{cb}$ 
and pressure $P_{cb}$ at the convective-radiative boundary are 
given by\footnote{These results coincide with equations
(47) and (48) of Rafikov (2006) if we identify 
$\nabla_{ph}=\nabla_\infty$ and assume $\nabla_{ph}\ll 1$.} 
\ba
&& T_{cb}=T_{ph}\left(\frac{\nabla_0-\nabla_{ph}}
{\nabla_0-\nabla_{ad}}\right)^{1/(4-\beta)},
\label{eq:T_cb}\\
&& P_{cb}=P_{ph}\left(\frac{\nabla_{ad}}{\nabla_{ph}}\cdot
\frac{\nabla_0-\nabla_{ph}}
{\nabla_0-\nabla_{ad}}\right)^{1/(1+\alpha)},
\label{eq:P_cb}
\ea
which can be derived by setting $\nabla(T_{cb})=\nabla_{ad}$
and using equation (\ref{eq:PofT}). Using equations 
(\ref{eq:hydro_eq}), (\ref{eq:PofT}), and (\ref{eq:P_cb}) 
one can also find that the optical depth at the 
convective-radiative boundary
\ba
\tau_{cb}\sim  \nabla_{ph}^{-1},
\label{eq:tau_cb}
\ea
while the radial extent of the outer radiative zone is 
\ba
\Delta R_r\sim H_{ph}\ln\nabla_{ph}^{-1}.
\label{eq:dR_r}
\ea
where $H_{ph}=k_B T_{ph}/(\mu g)$ is the photospheric scale 
height. Given that irradiation cannot heat 
the star to a temperature comparable to its central temperature
(otherwise outer layers would be unbound) $H_{ph}$ should be 
much smaller than $R_\star$ even under rather extreme irradiation.

Both $\nabla_0$ and $\nabla_{ad}$ are constants of order unity. 
This makes it clear from equation (\ref{eq:T_cb}) that the 
temperature variation between the photosphere and the convective
zone boundary is rather small, $|T_{cb}-T_{ph}|\sim T_{ph}$.
In practice, we find that at $T\lesssim 5000$ K, when $\kappa$
is given by (\ref{eq:low_T}), convection sets in at 
$T_{cb}\approx 1.36 T_{ph}$, while at higher temperatures, when 
$\kappa$ is given by (\ref{eq:high_T}), $T_{cb}\approx 1.19 T_{ph}$.
At the same time, under strong irradiation the external radiative 
zone should be deep enough for the pressure $P_{cb}$ at its bottom
to greatly exceed $P_{ph}$. In this case equation (\ref{eq:P_cb}) 
suggests that
\ba
\nabla_{ph}\ll 1,
\label{eq:nabla_cond}
\ea
a result that is verified in \S \ref{sect:isolated}, see 
equation (\ref{eq:nabl_ph}).

According to equation (\ref{eq:tau_cb}) smallness of 
$\nabla_{ph}$ results in $\tau_{cb}\gg 1$, thus justifying our
assumption (1) about the radiative zone properties.
At the same time, because of rather weak (logarithmic) dependence 
of $\Delta R_r$ on $\nabla_{ph}$, the thickness of the outer 
radiative zone should not be much different from $H_{ph}\ll 
R_\star$. As
a result, $\Delta R_r\ll R_\star$, verifying our assumption (2).
Thus, the condition (\ref{eq:nabla_cond}) can then be viewed as a 
prerequisite for the formation of a geometrically thin, optically thick  
radiative zone with roughly isothermal temperature profile 
under the action of intense external irradiation. External radiative 
layers with similar near-isothermal structure are expected to exist in
the envelopes of irradiated hot Jupiters (Guillot \etal 1996; Baraffe 
\etal 2003; Chabrier \etal 2004) and 
in the outer parts of the low-luminosity atmospheres of protoplanetary 
cores immersed into the protoplanetary nebulae (Rafikov 2006).

The value of $P_{ph}$ 
can be fixed in the following way. Above the photosphere gas is 
roughly isothermal with temperature $T_{ph}$ -- an 
approximation which is good enough for our purposes. Then at 
height $z$ above the photosphere
$\rho(z)=\rho_{ph}\exp(-z/H_{ph})$, where $\rho_{ph}$ is the 
photospheric gas density. Using this result and equation
(\ref{eq:kappa}) we find the photospheric optical depth
\ba
\frac{2}{3}=\int\limits_0^\infty\kappa\rho dz=
\frac{\kappa_{ph}P_{ph}}{(\alpha+1)g},
\label{eq:2_3}
\ea
from which it follows that
\ba
P_{ph}=\left[\frac{2(\alpha+1)}{3}\frac{g}{\tilde 
\kappa T_{ph}^\beta}\right]^{1/(1+\alpha)}.
\label{eq:P_ph}
\ea
As a byproduct of relation (\ref{eq:2_3}) one can rewrite 
equation (\ref{eq:nabla_ph}) as 
\ba
\nabla_{ph}=\frac{\alpha+1}{8}\frac{F_{in}}{\sigma T_{ph}^4}.
\label{eq:nab_ph}
\ea
It then follows from equations (\ref{eq:nabla_cond}) 
and (\ref{eq:nab_ph}) that $F_{in}\ll \sigma T_{ph}^4$.

We are now in position to evaluate $F_{in}$ and see how irradiation 
affects cooling of convective objects. To do this we note that
the inner boundary of the radiative zone is also the outer boundary
of the convective interior. We assume that convective 
transport is so efficient that entropy is constant\footnote{In 
reality stellar envelope contains superadiabatic regions which 
are not captured in our analysis and may affect its results.} 
throughout the inner convective zone, so that the EOS can be 
well represented by $P=K\rho^\gamma$, where $K$ is the adiabatic 
constant. As a result, $P_{ph}$ and $T_{ph}$ must be 
related via $(kT_{cb}/\mu)^\gamma=K P_{cb}^{\gamma-1}$ which, 
coupled with equations (\ref{eq:nabla_ph}), (\ref{eq:T_cb}), 
(\ref{eq:P_cb}), and condition (\ref{eq:nabla_cond}), yields
the following expression for $F_{in}$:
\ba
&& F_{in}(\theta)=\frac{16\nabla_{ad}}{3}
\left(\frac{\nabla_0-\nabla_{ad}}{\nabla_0}\right)^{\nabla_0/\nabla_{ad}-1}
\frac{\sigma g}{\tilde \kappa}\nonumber\\
&& \times\left(\frac{\mu K^{1/\gamma}}{k_B}\right)^{(1+\alpha)/\nabla_{ad}}
\left[T_{ph}(\theta)\right]^{4-\xi},
\label{eq:F_in}
\ea
where 
\ba
\xi=\beta+(1+\alpha)/\nabla_{ad}. 
\label{eq:xi}
\ea
Intrinsic stellar flux $F_{in}$ exhibits an explicit 
latitudinal dependence because it is a function of 
$T_{ph}(\theta)$.

As discussed before, when $\beta<4$ and $\nabla_0>0$ a transition to 
convection at some depth requires $\nabla_0>\nabla_{ad}$. 
As a result, 
\ba
4-\xi=(4-\beta)\left(1-\frac{\nabla_0}{\nabla_{ad}}\right)<0. 
\label{eq:4xi}
\ea
On the other hand, when 
$\beta > 4$ one also finds $4-\xi<0$ because $\nabla_0<0$ 
in this case. Thus, in both situations 
$F_{in}$ {\it decreases} as $T_{ph}$ increases. In other words,
irrespective of the opacity behavior external irradiation of the 
stellar surface {\it suppresses} 
stellar cooling, a result known from the studies of irradiated 
giant planets (Guillot \etal 1996; Burrows 2000).

Since external radiative zone is rather thin compared to $R_\star$ 
it must contain negligible amount of mass compared with $M_\star$. 
Then the structure of fully convective inner region of the 
star should be well described by the classical 
theory of polytropic spheres (Landau \& Lifshitz 1984; Kippenhahn 1994). 
In particular, adiabatic constant $K$ can be related to the stellar
mass and radius as
\ba
K=\zeta(\gamma) G M_\star^{2-\gamma}R_\star^{3\gamma-4},
\label{eq:K}
\ea
where $\zeta(\gamma)\sim 1$ is a parameter set by the equation 
of state of the gas. In a particular case of convective young stars 
with fully ionized interior characterized by $\gamma=5/3$ one 
has $\zeta(5/3)=0.1286$ and  
\ba
K=1.081\times 10^{14}~M_1^{1/3}R_{11}.
\label{eq:K_3_2}
\ea

Equations (\ref{eq:F_in}) and (\ref{eq:K}) 
unambiguously determine cooling 
of the star as a function of stellar parameters $R_\star$ and
$M_\star$, temperature distribution at the photosphere 
$T_{ph}(\theta)$, and opacity behavior in the outer radiative zone.


\subsection{Comparison with the case of an isolated star.}
\label{sect:isolated}

We now compare stellar cooling in the irradiated case with that 
occurring in isolated stars, in the absence of external 
illumination. In the latter case $T_{ph}=T_0$, $F_{in}=F_0=\sigma T_0^4$ 
and equation (\ref{eq:nab_ph}) gives 
$\nabla_{ph}=\nabla_{eff}=(\alpha+1)/8\sim 1$. Substituting this result 
into equations (\ref{eq:T_cb}), (\ref{eq:P_cb}), using 
adiabatic relation at the convective-radiative boundary, 
and equation (\ref{eq:P_ph}) we find
\ba
&& T_0=\left[\frac{16\nabla_{ad}}{3}
\left(\frac{\nabla_0-\nabla_{eff}}{\nabla_0-\nabla_{ad}}
\right)^{1-\nabla_0/\nabla_{ad}}\right.
\nonumber\\
&& \left.\times\frac{g}{\tilde \kappa}
\left(\frac{\mu K^{1/\gamma}}
{k_B}\right)^{(1+\alpha)/\nabla_{ad}}\right]^{1/\xi}.
\label{eq:T_0}
\ea
This expression sets
the effective temperature of the star and its cooling 
rate $F_0$ as functions of $M_\star$, $R_\star$ and
opacity behavior. In particular, for $\kappa$ typical 
at temperatures below $5000$ K one finds using equation 
(\ref{eq:K_3_2}) that
\ba
T_0\approx 1200~\mbox{K}~M_1^{11/39}R_{11}^{1/13}.
\label{eq:Hayashi}
\ea
This is considerably smaller than $T_{eff}\approx 3000-4000$ 
K typical for an isolated fully convective star on the Hayashi 
track that one obtains with detailed numerical stellar structure 
calculations (Siess \etal 2000). We ascribe this difference 
to our adoption of fixed $\gamma$ throughout the whole star and
the neglect of superadiabaticity in the outer parts of the 
convective region, see \S \ref{sect:complications}. At the same 
time equation (\ref{eq:Hayashi}) captures the 
main property of the Hayashi track -- extremely weak 
sensitivity of $T_0$ to $R_\star$ and, consequently, stellar 
luminosity. 

If we now go back to equation (\ref{eq:F_in}) one can easily
see that it can be rewritten as
\ba
F_{in}\approx F_0\left[\frac{T_{ph}(\theta)}{T_0}
\right]^{4-\xi},
\label{eq:Fin}
\ea
or, with equation (\ref{eq:balance}), as 
\ba
\left(\frac{F_{in}}{F_0}\right)^{4/(4-\xi)}=\frac{F_{in}}{F_0}
+\left(\frac{T_{irr}}{T_0}\right)^4.
\label{eq:Fin_alt}
\ea
This result together with (\ref{eq:4xi}) once again vividly 
illustrates the inhibition of stellar cooling by external 
irradiation and specifies the magnitude of this effect.

Using equations (\ref{eq:nab_ph}) and (\ref{eq:Fin}) we can
also write
\ba
\nabla_{ph}\approx \frac{(\alpha+1)}{8}\left(\frac{T_{ph}}{T_0}
\right)^{-\xi},
\label{eq:nabl_ph}
\ea
which shows that $\nabla_{ph}\ll 1$ when stellar
surface is strongly irradiated ($T_{ph}\gtrsim T_0$), thus confirming 
equation (\ref{eq:nabla_cond}). Note the
strong dependence of $\nabla_{ph}$ on $T_{ph}/T_0$: with our 
power-law anzatz for opacity $\xi\approx 13/2$ and $\approx 13$ 
below and above $5000$ K correspondingly, see equations (\ref{eq:low_T})
and (\ref{eq:high_T}).


\subsection{Conditions of validity of 1D approximation.}
\label{sect:1D_validity}

In Appendix \ref{ap:1D_validity} we determine the circumstances 
under which the results of \S \ref{sect:1D} hold true. We show
there that the condition of the validity of 1D approximation
can be expressed as
\ba
\left(\frac{H_{ph}}
{L_\theta}\right)^2\lesssim \nabla_{ph},
\label{eq:validity}
\ea
where $L_\theta$ is a characteristic scale in $\theta$ 
direction over which the external boundary condition 
[in our case $T_{ph}(\theta)$] experiences variation. Equation 
(\ref{eq:irr_flux}) and Figure \ref{fig:irr_flux} 
demonstrate that in the case of irradiation 
by accretion disk $L_\theta\sim R_\star$. Then equations 
(\ref{eq:nab_ph}) and (\ref{eq:Fin}) allow us to rewrite the 
condition (\ref{eq:validity}) as (assuming that $T_0$
and $T_{ph}$ are in the same opacity regime)
\ba
T_{ph}\lesssim T_0\left(\frac{R_\star}{H_0}
\right)^{2/(2+\xi)},
\label{eq:c1}
\ea
where $H_0=k_B T_0/(\mu g)$ is the photospheric scale height 
in the absence of irradiation. Given that
\ba
\frac{R_\star}{H_0}\approx 5\times 10^3~ M_1 R_{11}T_{3.5}^{-1}
\label{eq:rat}
\ea
we may conclude that 1D approximation should be rather accurate
even if $T_{ph}$ exceeds $T_0$ by a factor of several 
(e.g. $T_{ph}\lesssim 7 T_0$ for $T\lesssim 5000$ K). 

Whenever the condition (\ref{eq:validity}) is violated 
the redistribution of energy in 
$\theta$-direction within the radiative layer becomes  
important. In this case one needs to solve the full 
two-dimensional equation (\ref{eq:rad_tran}) without assuming
that radiative flux in $\theta$ direction is small. 
A similar situation arises at stellar equator where 
a lot of energy is released in a boundary layer 
that is not very extended in $\theta$ direction 
(Popham \etal 1993). As a result, at equator
$L_\theta\ll R_\star$ and the condition (\ref{eq:validity}) 
can be violated there even though at all other latitudes 
1D approximation works fine.


\section{Integrated stellar cooling.}
\label{sect:total_cooling}


We are now in position to calculate the integrated intrinsic 
luminosity $L$ (due to stellar contraction and interior cooling) 
of a convective star that is irradiated by a circumstellar disk:
\ba
L=4\pi R_\star^2\int\limits_0^{\pi/2}F_{in}(\theta)
\sin\theta d\theta,
\label{eq:L}
\ea
where $F_{in}(\theta)$ is given by the expression (\ref{eq:F_in}) 
in the irradiated part of the stellar surface, for 
$\theta\gtrsim\theta_{irr}$, while $F_{in}(\theta)\approx 
\sigma T_0^4$ in the weakly irradiated polar regions, for
$\theta\lesssim\theta_{irr}$. 

Convective objects can exhibit different modes of cooling which  
is best illustrated by considering the limit $\theta_{irr}\ll 1$
($\Lambda\gg 1$). In this limit the contribution of polar
caps to the total luminosity is
\ba
L_{pc}\approx 4\pi F_0 R_\star^2(1-\cos\theta_{irr})\approx 
4\pi F_0 R_\star^2 \theta_{irr}^2,
\label{eq:L_pc}
\ea
while irradiated equatorial regions contribute
\ba
L_{er}\approx \frac{4\pi C F_0 R_\star^2} 
{[I\sin^5\theta_{irr}]^{(4-\xi)/4}}\int
\limits_{\theta_{irr}}^{\pi/2}[g(\theta)]^{(4-\xi)/4}
\sin\theta d\theta,
\label{eq:L_er}
\ea
see equation (\ref{eq:Fin}).

\begin{figure}
\plotone{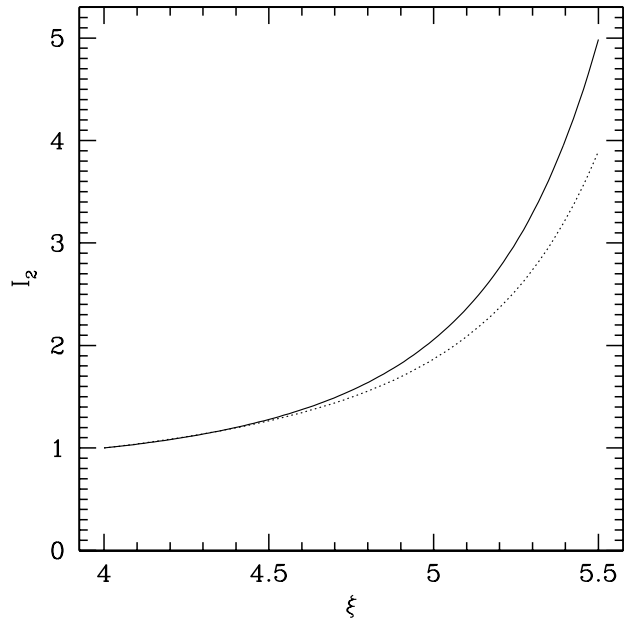}
\caption{
Plot of function $I_2(\xi)$. Solid line corresponds to $I_2$
computed using $g(\theta)$ (see Figure \ref{fig:irr_flux}) and 
the dashed line is $I_2$ calculated
using asymptotic representation (\ref{eq:as}).
\label{fig:I_2}}
\end{figure}

Using equations (\ref{eq:as}), (\ref{eq:theta_c}) 
it is easy to see that for 
$\theta_{irr}\ll 1$ the latter integral is dominated by 
$\theta\approx \theta_{irr}$ if 
\ba
\xi>\frac{28}{5}.
\label{eq:xi_cond}
\ea
In this case, according to equation (\ref{eq:as}), 
one may approximate $g(\theta)\approx I\sin^5\theta$ and
find that 
\ba
L_{er}\approx \frac{16\pi}{5(\xi-28/5)} 
F_0 R_\star^2 \theta_{irr}^2\sim L_{pc}. 
\label{eq:L_er1}
\ea
This results leads us to an interesting conclusion that as long
as the condition (\ref{eq:xi_cond}) is fulfilled, an object
cools predominantly through its polar caps and its 
integrated luminosity $L$ is almost independent of the details 
of opacity behavior in its outer layers. The latter point 
is easy to understand, since in this case $L\sim F_0 S$, 
where $S$ is the
surface area of the polar caps. But according to equation 
(\ref{eq:theta_c}) the value of $S$ is determined only by 
irradiation and $T_0$. As a result, $L$ depends on $\kappa$ 
only weakly, through the $L_{er}$ contribution. 

We call the regime of stellar cooling realized under 
the condition (\ref{eq:xi_cond}) the {\it high-latitude} 
cooling. This regime naturally occurs in irradiated 
young stars since $\xi>28/5$ for $\kappa$ given by either
(\ref{eq:low_T}) or (\ref{eq:high_T}). Equations (\ref{eq:L_pc})
and (\ref{eq:L_er1}) demonstrate that in this regime $L$ 
is suppressed roughly by $\sim \theta^2_{irr}$
which may be as low as $\sim 0.2-0.4$ according to the 
expression (\ref{eq:theta_c}). Thus, disk irradiation 
can substantially slow down cooling of
young stars.

In the opposite case of $\xi<28/5$ cooling is in the 
{\it low-latitude} regime, so that star loses most
of its internal energy through the equatorial regions 
even though they are strongly irradiated. In this case
one should use the full expression (\ref{eq:L_er}) to 
evaluate $L\approx L_{er}$. Stellar luminosity suppression 
for $\theta_{irr}\ll 1$ is given by 
\ba
&& L/L_0\approx C I_2(\sin\theta_{irr})^{5(\xi-4)/4},
\label{eq:second_as}\\
&& I_2(\xi)=I_1^{(\xi-4)/4}\int
\limits_{0}^{\pi/2}[g(\theta)]^{(4-\xi)/4}
\sin\theta d\theta.\nonumber
\ea
Function $I_2(\xi)\sim 1$ is shown in Figure \ref{fig:I_2}.
Knowing that $4<\xi<28/5$ in the low-latitude case one 
can easily see that the degree of luminosity 
suppression is smaller than in the high-latitude 
cooling regime.

\begin{figure}[t]
\plotone{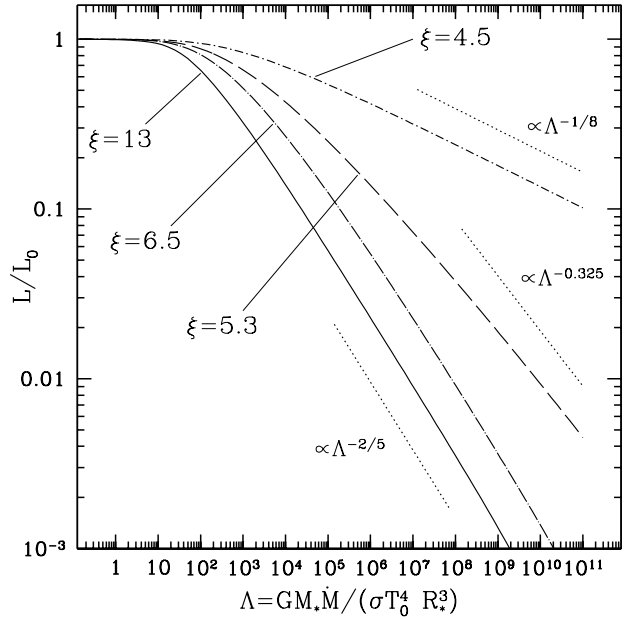}
\caption{
Plots of the intrinsic luminosity of an irradiated star L 
(in units of an isolated star luminosity $L_0$) as a function 
of irradiation parameter 
$\Lambda=GM_\star \dot M/(\sigma T_0^4 R_\star^3)$ for 
different values of the power law index $\xi$ defined by equation
(\ref{eq:xi}). Dotted lines illustrate the corresponding asymptotic 
behaviors for $\Lambda\gg 1$: $L/L_0\propto \Lambda^{-2/5}$
for $\xi>28/5$ and $L/L_0\propto \Lambda^{(4-\xi)/4}$
for $\xi<28/5$. 
\label{fig:supp}}
\end{figure}

In Figure \ref{fig:supp} we plot $L/L_0$ -- the ratio of 
stellar luminosities in the irradiated and isolated cases 
-- as a function of irradiation parameter $\Lambda$, 
for different values of $\zeta$. This calculation does not 
explicitly make an assumption $\theta_{irr}\ll 1$ (or 
$\Lambda\gg 1$), although it 
covers this regime as well. Here $L/L_0$ is computed by the 
straightforward integration of $F_{in}$ over the stellar 
surface (including the polar caps where $F_{in}=F_0$), with
the distribution of $T_{irr}(\theta)$ found from equation
(\ref{eq:irr_flux_mod}) and $g(\theta)$ displayed in Figure 
\ref{fig:irr_flux}. We also indicate the asymptotic
behavior of $L/L_0$ as given by equations (\ref{eq:L_pc}) 
and (\ref{eq:L_er1}) for $\xi>28/5$ and equation 
(\ref{eq:second_as}) for $\xi<28/5$. 

One can easily see that, 
as expected, $L/L_0\propto\theta_{irr}^2\propto \Lambda^{-2/5}$
as $\Lambda\gg 1$
for $\xi=13$ and $6.5$ independent of the actual value of $\xi$
(only the normalizations of the curves are different because of the
different contributions produced by the near-polar cap regions) since 
for both $\xi>28/5$. Significant suppression of the stellar 
flux (by $\sim 2$) is found in this case already at 
$\Lambda\sim 10^2-10^3$. In the case $\xi<28/5$ asymptotic behavior
for $\Lambda\gg 1$ agrees well with equation (\ref{eq:second_as}),
$L/L_0\propto \Lambda^{(4-\xi)/4}$, and
the degree of stellar flux suppression is weaker than in the 
high-latitude regime: $L/L_0\approx 0.5$ only at 
$\Lambda\approx 5\times 10^3$ for $\xi=5.3$
and at $\Lambda\approx 2\times 10^5$ for $\xi=4.5$.

Note that results presented in Figure \ref{fig:I_2} are calculated
neglecting any additional heating that can be produced near the 
equator by the boundary layer dissipation. We will address this 
point in more detail in \S \ref{sect:complications}.

\section{Discussion.}
\label{sect:disc}


Luminosity suppression by disk irradiation is one of the most important 
results of this work. Analogous phenomenon has been previously 
found in studies of extrasolar giant planets in short-period orbits, 
where stellar irradiation is quite severe (Guillot \etal 1996; 
Burrows \etal 2000). In that case irradiation affects only one 
side of the plant which always faces the star. Some heat from
the day side gets redistributed to the night side by atmospheric
circulation (Menou \etal 2003; Dobbs-Dixon \& Lin 2007) 
which should complicate the calculation of the 
photospheric boundary conditions across the whole planetary 
surface. In our case irradiation is azimuthally symmetric which
makes our calculation more robust. Analogous to the case of 
extrasolar giant planets we expect that luminosity 
suppression by irradiation would tend to retain heat inside the star 
and increase stellar radius above the value found in the absence of 
irradiation (Baraffe \etal 2003). Whether this radius increase is 
significant will be investigated in future work.

Luminosity suppression may have some effect on the strength of the 
magnetic field that is generated by dynamo action in the convective 
interior of the star. Since in irradiated case convective eddies 
transport smaller energy flux to the stellar surface than in the 
case of an isolated star the speed of convective motions is expected 
to be smaller. This results in a less vigorous dynamo action and 
likely weaker magnetic field generated inside the star. 

Formation of an optically thick radiative zone near the stellar surface
in irradiated regions is another result of this work which has
important implications.  As the star grows the hot gas in the vicinity 
of the boundary layer where disk meets the star 
gets advected into the convective interior thereby raising stellar 
entropy. This provides another way of slowing down stellar contraction,
in addition to the luminosity suppression discussed above. Hartmann 
\etal (1997) argued that impact of the heat advection on stellar structure 
is not significant as long as the temperature of 
advected gas is much smaller than the central temperature of the star.
In the irradiated case it is the properties of the external radiative 
zone that determine the temperature of the gas at the 
convective-radiative boundary, and thus the amount of thermal energy 
advected into the stellar interior. Indeed, hot gas sinking through the 
radiative zone will lose a significant fraction of its thermal 
energy by radiative diffusion in the latitudinal direction, so that 
the temperature at the convective-radiative boundary is likely to be 
lower than in the center of the boundary layer. It is important to build the 
detailed 2D model of the radiative transport in the vicinity of the 
boundary layer to quantify this effect and to verify the significance of 
heat advection (see also \S \ref{sect:1D_validity}).

Presence of the radiative zone may also affect atmospheric opacity in 
accreting brown dwarfs and giant planets. Accreted gas brings in
significant amount of dust into the object's atmosphere (dust can 
also form out of the gas phase under low-temperature conditions) 
which changes $\kappa$ and radiative properties of the star 
(Chabrier \etal 2000). However, if the object is fully 
convective all the way to its photosphere, vertical fluid motions
quickly advect dust grains into the hot interior where grains get 
easily destroyed. This is not the case in irradiated regions of accreting 
objects since dust grains can hover in the radiative zone for a 
long time, as long as their gravitational settling is not too fast. 
Of course, for grains to exist in the outer radiative layer  
in the first place $T_{irr}$ must be lower than the sublimation 
temperature of dust material, which may be possible only in accreting 
brown dwarfs and giant planets. 

We expect that in the case of young stars it would be very difficult 
to obtain a direct observational confirmation of the luminosity 
suppression by irradiation. The major reason for this 
is that $L/L_0$ starts to deviate from unity only when $\dot M$ and,
correspondingly, accretion luminosity are very large. At this stage 
the luminosity of a star+disk system is completely dominated by 
the direct emission from the disk and the disk flux intercepted 
and re-radiated by the stellar surface. Intrinsic stellar luminosity 
provides negligible contribution which would be almost impossible
to distinguish. Besides, forming protostar should still be enshrouded 
in the dense veil of the residual gas collapsing onto the 
circumstellar disk. Reprocessing of star+disk emission in
this infalling envelope would complicate things even more. 
Another potential way of detecting the luminosity suppression 
is indirect, through its effect on the stellar radius and luminosity 
as the star emerges as almost fully formed Class I object in the end 
of active accretion phase. 

Effects of disk accretion in star formation have been previously 
investigated by a number of authors. Adams \& Shu (1986) and 
Popham (1997) have calculated the amount of energy which is 
emitted by the disk and is intercepted by the star. Unlike us 
these authors were not primarily concerned in the details of 
the distribution of irradiation flux over the stellar surface. 
This is a crucial point of our study allowing us to identify 
the two different regimes of stellar cooling -- high- and 
low-latitude. 

Mercer-Smith \etal (1984) were the first to explore the effect 
of disk accretion on the stellar structure. 
They handled disk accretion by specifying mass addition 
rate and accretion luminosity as external boundary conditions.
They find that the stars formed by disk accretion have {\it larger}
radii than nonaccreting stars of the same mass. As mentioned in 
Hartmann \etal (1997) this outcome most likely results from allowing
the accreted material to have very high entropy which leads to 
stellar swelling, see Prialnik \& Livio (1985). This approximation 
is unlikely to be valid in reality since disk 
material joining the stellar surface should have enough time to 
radiate most of its thermal energy before being fully incorporated 
into the star. Palla \& Stahler (1992) and Hartmann \etal (1997) 
in their studies of intermediate- and low-mass stars allowed the 
accreted material to have low entropy. They found that accretion 
{\it reduces} stellar size compared to the non-accreting case since 
in this case the addition of mass leads only to 
the increase of gravitational energy of the star and is not accompanied by
the increase of thermal energy. 

All these studies have either ignored irradiation of star by
the disk or accounted for it only in the averaged sense which may
not be acceptable as our study demonstrates. To get a complete 
picture of protostellar evolution one needs to include the 
luminosity suppression by disk irradiation into account. 
Such calculation must necessarily 
allow for the spatial distribution of irradiation flux on the 
stellar surface since only in this way a proper estimate of
the luminosity suppression can be obtained.


\subsection{Applications to real systems.}
\label{sect:applications}

Here we apply our results to different classes of fully convective 
objects which may accrete through the disk at high $\dot M$. In doing
our estimates, which require the knowledge of $R_\star$ and $T_0$, 
we use the radii and photospheric temperatures of 
corresponding objects determined in the absence of irradiation and mass 
inflow by accretion as proxies for  $R_\star$ and $T_0$ that 
these objects would have if irradiation and accretion were properly 
accounted for. Needless to say, a truly accurate estimate of the 
effect of irradiation can be obtained only if $R_\star$ and 
$T_0$ are calculated self-consistently accounting for the 
effects of accretion and irradiation.

\subsubsection{Young stars}
\label{sect:youngstars}

\begin{figure}[b]
\plotone{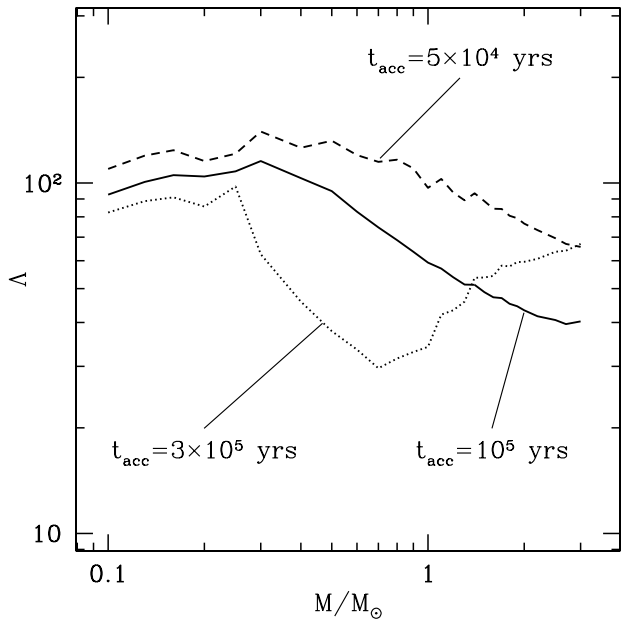}
\caption{
Irradiation parameter $\Lambda$ for stars of different mass
and age $t_{acc}$ (indicated on the plot) assembled by disk accretion
with constant $\dot M=M_\star/t_{acc}$. Stellar parameters from
Siess \etal (2000) were used in this calculation.
\label{fig:lambdas}}
\end{figure}

Young low-mass stars transitioning from Class 0 to Class I 
phase are fully convective and should be 
assembled by mass accretion from a circumstellar disk within 
several $10^5$ yrs. This implies very high accretion rate 
and we adopt $\dot M=5\times 10^{-6}$ M$_\odot$ yr$^{-1}$ for a 
simple  estimate.  An isolated $M_\star=0.5$ M$_\odot$ star 
at an age of $10^5$ yrs has a radius $R_\star=3.9$ R$_\odot$ and 
effective temperature $T_0=3760$ K (Siess \etal 2000). 
These parameters yield $\Lambda=10^2$ and $T_{irr}(\pi/2)
\approx 6000$ K, although the latter is likely to be higher 
because of the boundary layer dissipation. At this $\Lambda$ 
cooling in the equatorial region is suppressed but the size 
of the polar caps is reduced only weakly: according to Figure 
\ref{fig:irr_flux} $\theta_{irr}\approx 73^\circ$ 
[$\theta_{irr}$ is given by an implicit 
relation $g(\theta_{irr})=\Lambda^{-1}$]. In Figure 
\ref{fig:lambdas} we present more general results for $\Lambda$ 
calculated for stars of different masses using stellar
parameters from Siess \etal (2000) and assuming 
constant $\dot M=M_\star/t_{acc}$, where $t_{acc}$ is the accretion 
time (assumed equal to the stellar age). 
One can see that the low-mass stars assembled within
$3\times 10^5$ yrs generally have $\Lambda$ in the range of
$30-10^2$, agreeing with our simple estimate.

Since $\xi\approx 6$ for $T\lesssim 5000$ K, young stars cool 
mainly through the polar regions and we find from Figure
\ref{fig:supp} that stellar luminosity is reduced by 
irradiation only by about $10\%$ for 
$\Lambda\approx 10^2$. On the other hand, if $\dot M$
is not constant but increases as $M_\star$ grows one may expect
values of $\Lambda$ larger by a factor of several. Also, to 
calculate $\Lambda$ we have used parameters of isolated stars 
while stars assembled by disk accretion of the low-entropy material 
have smaller $R_\star$,
leading to larger $\Lambda$. All these factors may increase the
importance of disk irradiation in determining $L$ of 
young accreting stars.

\subsubsection{Young stars in quasar disks}
\label{sect:youngstars_QSOs}

A very interesting mode of star formation is possible in the 
accretion disks around the supermassive black holes (SMBHs) in the
centers of galaxies (Illarionov \& Romanova 1988; 
Goodman \& Tan 2004; Nayakshin 2006).
It is currently known that our own Galactic Center harboring 
a SMBH of mass $M_{BH}\approx 3.7\times 10^6$ M$_\odot$ 
(Ghez \etal 2005) contains 
a number of young ($\lesssim 6$ Myrs) massive 
($M_\star\gtrsim 10$ M$_\odot$) stars that form 
two misaligned disks around the SMBH (Paumard \etal 2006).
One of the most likely scenarios for the origin of these stars 
is a fragmentation of a gravitationally unstable gaseous disk 
(or disks) followed by the growth of fragments to their present
masses by gas accretion from the residual disk (Levin 2006; 
Nayakshin 2006). Assuming that disk temperature is kept at 
the level of $50$ K by the radiation of nearby stars (Levin 2007) 
one finds that at $a=0.1$ pc from the SMBH 
(which is the typical dimension of the observed disks)  
surface mass density of $\Sigma\approx 27$ g cm$^{-2}$ is 
required for the disk to be Toomre unstable. 

Fragments formed as a result of instability at $0.1$ pc 
have a typical mass
$M_\star\sim \Sigma h^2\approx 10^{-3}$ M$_\odot$ (approximately one 
Jupiter mass) where $h$ is a disk scale height. At formation
the Hill radius of such an object $R_H=a(M_\star/M_{BH})^{1/3}$
is already comparable to $h$ and as $M_\star$
grows by accretion $R_H$ becomes larger than $h$. As a result,
accretion onto fragment proceeds through the {\it sub-disk} that 
forms within the fragment's Hill sphere, presenting us with the
setting investigated in this paper. Rate at which gas flows into
the fragment's Hill sphere is the Hill accretion rate\footnote{Such 
high $\dot M$ is also typical for FU Orioni objects. As demonstrated by 
Popham \etal (1993) in the high-$\dot M$ regime the boundary layer 
is so thick that it covers a significant ($\sim 0.5$) fraction of 
the stellar surface slowing down interior cooling. Heat advection 
into the stellar interior may also become an issue (Popham 1997).} 
$\dot M_H\approx 
\Sigma\Omega R_H^2\approx 2\times 10^{-4}~M_1^{2/3}(a/0.1~\mbox{pc})^2$ 
M$_\odot$ yr$^{-1}$. Note that $\dot M_H$ is smaller
than the Eddington mass accretion rate 
$\dot M_{Edd}=4\pi cR_\star/\kappa_{es}=
1.4\times 10^{-3}R_{11}$ M$_\odot$ yr$^{-1}$ (here $c$ is the 
speed of light and $\kappa_{es}$ is the electron scattering opacity) at 
$a=0.1$ pc but may become comparable to $\dot M_{Edd}$ further out from 
the SMBH provided that $R_\star$ is not much larger than $R_\odot$.
 
If gas in the disk is able to accrete at the  
same high rate $\dot M_H$ onto the stellar surface then 
\ba
\Lambda\approx 3\times 10^5 M_1^{5/3}T_{3.5}^{-4}
R_{11}^{-3}.
\label{eq:Hill_Lambda}
\ea
At present we do not have a theory for the structure of 
stars formed by fragmentation of gravitationally unstable disks
so the value of $R_\star$ is highly uncertain. If $R_\star \lesssim 10~R_\odot$ 
then $\Lambda\gtrsim 10^3$ and luminosity suppression by irradiation 
should be quite important,
reducing $L$ by a factor of $2-3$ compared to $L_0$, as Figure \ref{fig:supp} 
demonstrates for $\xi=6.5$. As the value of $R_\star$ itself is 
affected by the time history of $L$, irradiation
should not be overlooked in studies of star formation in quasar 
disks.

\subsubsection{Young brown dwarfs}
\label{sect:BDs}

Brown dwarf (BD) formation is likely to be a scaled down version of
the low-mass star formation: one again expects a formation 
of a centrifugally supported disk around a fully convective
object that grows by disk accretion. The biggest uncertainty in 
determining $\Lambda$ is again $R_\star$: $0.1$ Gyr old  
BDs have radii of $0.1-0.2$ R$_\odot$ (Baraffe \etal 2003) 
but accumulation of their mass 
(poorly investigated at present) likely takes less than 
$10^5$ yr, during which time their entropy is still
quite high,
resulting in considerably larger $R_\star$. Assuming
that an object with $M_\star=0.03$ M$_\odot$ grows at constant 
$\dot M$ in time $t_{acc}=5\times 10^4$ yr and has $R_\star=0.5$ 
R$_\odot$ and $T_0\approx 3000$ K we find $\Lambda\approx 800$. 
Provided that opacity can still be characterized by expression 
(\ref{eq:low_T}) (which is a somewhat questionable assumption)
we conclude that irradiation may lead to order unity reduction 
in $L$. As the brown dwarf cools and contracts $\Lambda$ increases 
bringing down $L/L_0$ even more, provided that $\dot M$ could still be 
maintained at high level. Thus, young BDs may be affected by
the disk irradiation which may have consequences 
for their subsequent thermal evolution.

\subsubsection{Young giant planets}
\label{sect:planets}

Finally, we consider the situation arising during the late stages 
of giant planet formation via the so-called core instability. This
scenario of planet formation assumes buildup of a 
$\sim 10$ M$_\oplus$ refractory core in the protoplanetary nebula by
planetesimal agglomeration. The self-gravity of the core triggers an 
instability and leads to rapid gas accumulation (Mizuno 1980). 
While the initial stages of this process can be adequately described 
in the spherically-symmetric approximation, the later epoch of unstable
gas accretion must have distinctly non-spherical morphology. Indeed, as
mentioned in Rafikov (2006), as soon as the mass of a rapidly growing planet
exceeds the so-called {\it transitional} mass $M_{tr}=c_s^3/\Omega G\approx 
40 M_\oplus a_5^{3/4}$ (here $c_s$ is the gas sound speed in the nebula
and $a_5\equiv a/5$ AU is the planetary semi-major axis scaled by 5 AU) 
the Hill radius of the planet $R_H$ becomes larger than the scale 
height of the disk $h$. As a result, protoplanet starts accreting 
gas from the surrounding nebula through the sub-disk that forms 
within its Hill sphere, thereby presenting a situation analogous to 
the star formation in the Galactic Center described in 
\S \ref{sect:youngstars_QSOs} (except that now the collapsing 
fragment of a gravitationally unstable disk is replaced by a growing 
planet). Here we assess how important can irradiation be for planetary
cooling when $M_p\gtrsim M_{tr}$.

The maximum $\dot M$ available to the planet is still likely given 
by the Hill rate $\dot M_H\approx 2\times 10^{-3}M_{p,2}^{2/3}a_5^{-1}$ 
M$_{\rm J}$ yr$^{-1}$, where $M_{p,2}\equiv M_p/10^2$ M$_\oplus$ and
we have adopted a surface density profile $\Sigma=270 a_5^{-3/2}$ 
g cm$^{-2}$ typical for the Minimum-Mass Solar Nebula. This allows us
to compute
\ba
\Lambda\approx 10^4 M_2^{5/3}a_5^{-1}
\left(\frac{R_p}{5~\mbox{R}_J}\right)^{-3}
\left(\frac{T_0}{10^3~\mbox{K}}\right)^{-4}.
\label{eq:Lam_planet}
\ea
This estimate is rather uncertain because of poorly constrained
$R_p$ and $T_0$ during the stage of active gas accretion by the 
planet. Here we adopt $R_p=5$ R$_J$ and $T_0=10^3$ K mainly for 
illustrative purposes.

Dust is the major source of opacity in the outer layers of forming
giant planets. Dust opacity scales as $\kappa\propto T^{\beta}$
with $\beta\approx 0.5-2$ depending on dust grain composition,
spectrum of grain sizes, etc. Here we adopt $\beta=1$ in which
case $\xi=4.5$. This corresponds to cooling dominated by the 
equatorial regions, which is different from the stellar case. As a 
result, $L/L_0$ should be more sensitive 
to the structure of the boundary layer through which disk material 
accretes onto the planet, namely, $L/L_0$ should be {\it lower} 
than Figure 
\ref{eq:Lam_planet} implies.  Forgetting about this complication 
for the moment and using $\xi=4.5$ and $\Lambda$ from
equation (\ref{eq:Lam_planet}), we find from Figure \ref{fig:supp} 
that in the planetary case $L$ can be suppressed compared to 
$L_0$ by several tens of per cent. Given the existing uncertainties 
in modeling the late stages of planet formation this degree of 
luminosity suppression by irradiation may not seem like a 
serious issue. 
However, more massive, compact and cooler planets can easily have 
$\Lambda\sim 10^5-10^6$ in which case irradiation would reduce  
$L$ by a factor of several potentially affecting planetary evolution.


\subsection{Additional complications.}
\label{sect:complications}

Here we address various complications that may arise when the
results of this work are applied to real objects.

All our derivations and estimates explicitly assumed opacity 
in the form (\ref{eq:kappa}). While 
this representation can be quite accurate within some temperature 
intervals one has to bear in mind that on the surface of a star 
irradiated by the disk temperature can vary appreciably 
between the equator and the poles. Indeed, the equatorial 
regions of a $M_\star=1$ M$_\odot$, $R_\star=2$ R$_\odot$ star 
accreting at $\dot M=10^{-5}$ M$_\odot$ yr$^{-1}$ are heated
to $1.1\times 10^4$ K so that equation (\ref{eq:high_T}) applies, 
while polar caps still have $T\approx 3\times 10^3$ K so 
that equation (\ref{eq:low_T}) is more appropriate. Thus, in
different regions of stellar surface $\kappa$ has
different dependence on $P$ and $T$. In this case equation 
(\ref{eq:second_as})  becomes invalid and to properly 
compute the luminosity suppression one would need to take into account
the latitudinal variation of not only $T_{irr}$ but also  
$\kappa(P,T)$. 

Even at a fixed latitude opacity can switch from one 
regime to another within the outer radiative zone. Although $\kappa$ 
is much more sensitive to $T$ than to $P$ and the radiative zone 
is roughly isothermal, pressure at its bottom  
$P_{cb}$ is $\sim \nabla_{ph}^{-1/(1+\alpha)}\gg 1$ of the
photospheric pressure $P_{ph}$, so that even a weak dependence 
of $\kappa$ on $P$  can lead to opacity transition within the outer
radiative zone. In particular, this situation is likely to occur 
at latitudes where $T_{irr}\approx 5000$ K and $\kappa$ switches 
from (\ref{eq:low_T}) to (\ref{eq:high_T}). In this case  
calculation of the local radiative flux $F_{in}$ gets more complicated 
as the external radiative region splits into two layers characterized 
by different opacity behaviors. 

Our numerical estimates of the luminosity suppression strongly
rely on the assumption of fixed $\gamma$ within the radiative 
zone. Equation (\ref{eq:Hayashi}) which assumes $\gamma=5/3$ 
throughout the whole star fails to predict the correct photospheric
temperature of an isolated star. The reason for
this is the variation of $\gamma$ at low $T$
caused by molecular dissociation.
This results in $\nabla_{ad}\lesssim 0.1$ and leads to a 
smaller drop of temperature in the outer convective parts of 
the star. On the other hand, superadiabaticity of convection 
in the outer layers of convective zone counteracts this effect 
to some extent. Our
simple estimate (\ref{eq:Hayashi}) underpredicts $T_0$ by a factor 
of $2-3$ which is primarily a consequence of our assumption of fixed 
$\gamma=5/3$. A proper calculation of $F_{in}(\theta)$ and 
$L/L_0$ must be able to account for the variation of $\gamma$
with $P$ and $T$ inside the outer radiative zone.

Our analysis is affected to some extent by the presence of 
the boundary layer through which disk material 
joins the star. Viscous dissipation in this 
layer heats accreting gas to very high temperature. Since 
this energy release takes place very close to the stellar surface 
most of the heat is likely to leak out and not get carried into 
the star with the accreted gas. However, some residual 
heat may still be accreted. Moreover, in addition to advective there could also 
be a radiative energy transfer from the boundary layer into the outer
layers of the star  (Popham 1997). Increase of $T$ in the 
external radiative zone driven by these processes acts to 
additionally slow down stellar cooling in the equatorial region, as 
equation (\ref{eq:F_in}) demonstrates. Thus, presence of the 
boundary layer reduces $L/L_0$ even more than our analysis 
predicts, and the results presented in Figure \ref{fig:supp} 
should be viewed as upper limits on $L/L_0$. This effect is
likely not very important for young stars which cool predominantly
through their polar regions, largely unaffected by the 
additional heat deposition at
the equator. However, in the case of young giant planets which lose
fair amount of energy through the low-latitude part of the surface 
(see \S \ref{sect:planets}) the reduction of intrinsic flux in 
the equatorial region may produce quite noticeable decrease of 
$L/L_0$ compared to the idealized case considered in this work.

\begin{figure}[t]
\plotone{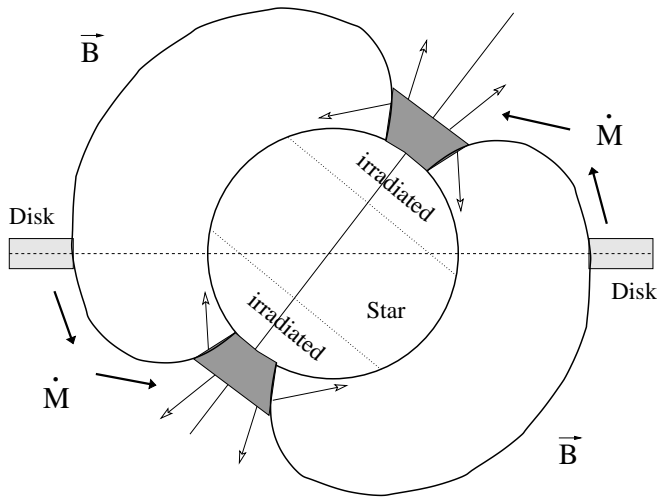}
\caption{
Schematic representation of magnetically channeled accretion. 
Disk material travels along the magnetic field lines and 
sediments onto the magnetospheric columns (shaded) heated by shock 
dissipation and gravitational settling of material in them. 
Hot magnetospheric columns irradiate stellar surface near the 
magnetic poles.
\label{fig:mag_pole}}
\end{figure}

When discussing the external radiative zone throughout this 
work we have been concerned only with the radiative energy
transport. At the same time, it is well known that in the case 
of hot Jupiters advective transport in the form of
atmospheric jets and winds can be quite important in
redistributing heat across the planetary surface (Menou 
\etal 2003; Dobbs-Dixon \& Lin 2007). 
In our azimuthally-symmetric setting only 
meridional atmospheric motions can lead to energy exchange 
between the hot equatorial and cold polar regions. Fluid 
motions occurring on surfaces of constant effective potential
(gravitational plus centrifugal) are unlikely to produce
efficient equator-pole energy exchange (compared to the 
radiative transfer) because of rather 
fast rotation typical for objects formed by disk accretion. Rotation
forces angular momentum conservation and prevents significant 
fluid motions in $\theta$-direction, suppressing this mode of
advective transport. On the other hand,  rotation tends
to promote meridional circulation within the radiative layer 
(Kippenhahn \& Weigert 1994) whose impact on the energy 
transport in the outer radiative zone should be 
investigated in more detail.

Finally, our basic assumption of direct mass accretion from 
the disk may be challenged if the growing star possesses magnetic 
field strong enough to disrupt accretion flow outside of 
$R_\star$ (K\"onigl 1991; Matt \& Pudritz 2005).
In this case gas is channeled by the magnetic field and 
is deposition onto the stellar surface at magnetic 
poles rather than at stellar equator. This completely changes 
the topology of accretion flow but the major conclusions about
the effect on stellar cooling are likely to hold. 
Indeed, the magnetically channeled gas travels towards 
the stellar surface at a good fraction of the free-fall velocity
and at some point it must pass through the radiative shock, after 
which it accumulates at the top of the magnetospheric column 
of accreted material, as schematically indicated in Figure 
\ref{fig:mag_pole}. Total energy release within the shock 
and magnetospheric column is comparable to that occurring 
if the accretion disk were extending all the way to the stellar 
surface. This hot column of accreted material illuminates the 
surface of the star leading to the same suppression of intrinsic 
stellar flux as we discussed in this work. In this case, however, 
irradiation is strongest near the magnetic poles while 
the magnetic equator is likely to be the coolest part of the
stellar surface\footnote{Illumination of the star by the distant parts of 
accretion disk, beyond the point where accretion flow is 
disrupted by the magnetic field, is 
unlikely to be very important given the rapid 
fall-off of $F_d$ with the distance from the star.}.  
Calculation of stellar irradiation and integrated luminosity in 
this case would involve constructing a model for the 
magnetospheric column structure and its radiative properties. 

The impact of these details on the structure and evolution 
of young stars should be addressed by future work. 


\section{Summary.}
\label{sect:concl}


Luminosity of young stars actively accreting from the 
circumstellar disk can be significantly affected by the 
radiation which is produced in the inner parts of the disk 
and is intercepted by the stellar surface. We showed that
if a star gains its mass via disk accretion on timescale 
of several $10^5$ yr then the radiative flux caused
by viscous dissipation in the disk is more than sufficient 
to increase the surface temperature of the star above the
photospheric temperature that an isolated star with the 
same mass and radius would have. Irradiation
by the disk is strongest in the equatorial regions and is
almost negligible near the poles. An outer radiative zone 
of almost constant temperature forms above the fully convective 
interior in the strongly irradiated parts of the stellar 
surface. This leads to the local suppression of intrinsic
energy flux escaping from the stellar interior. 

We have 
demonstrated that there are two distinct modes in which a
fully convective object can cool: mainly through the 
cool high-latitude polar regions or predominantly through the 
low-latitude parts of the stellar surface. A particular 
regime of cooling in a given object is set by the opacity 
behavior and the adiabatic temperature gradient $\nabla_{ad}$ in the outer 
radiative zone. Accreting young stars and brown dwarfs cool 
mainly through the polar regions while forming giant planets
cool through the whole surface.

Integrated stellar luminosity in accreting case is suppressed 
compared to the case of an isolated object, by up to a factor
of several in some classes of objects (actively accreting brown dwarfs
and planets, stars forming in gravitationally unstable 
disks in galactic nuclei). This leads to larger radii of 
irradiated objects and may affect the initial conditions which 
are used to calculate the evolution of the low-mass objects 
on timescales of $\sim 10$ Myr after their formation.
Existence of external radiative zone may facilitate retention 
of dust in the atmospheres of brown dwarfs and planets, and
may affect the strength of magnetic field generated by 
internal dynamo in convective objects.

Some of the results obtained in this work may be 
applicable to accreting white dwarf and neutron star systems.

\acknowledgements 

I am grateful to Gilles Chabrier for careful reading of the 
manuscript and many useful suggestion. 
The financial support for this work is provided
by the Canada Research Chairs program and a NSERC 
Discovery grant.

\appendix


\section{Irradiation flux.}
\label{ap:irr_flux}

Performing an integral over $\phi$ in eq. 
(\ref{eq:irr_flux}) we find
\ba
&& F_{irr}(\theta)=\frac{2\cos\theta}{\pi}
\int\limits_{x_{in}}^\infty \frac{F_d(xR_\star)q(x,\theta)x dx}{D^2},
\label{eq:1D}\\
&& q(x,\theta)=\sqrt{x^2\sin^2\theta-1}-2\frac{x^2(1-2\sin^2\theta)+1}{D}
\arctan\left(\frac{x^2+2x\sin\theta+1}{x\sin\theta+1}
\sqrt{\frac{x^2\sin^2\theta-1}{D^2}}\right),
\label{eq:h}
\ea
where $D^2=(x^2+1)^2-4x^2\sin^2\theta$ and $x_{in}=1/\sin\theta$.
For $F_d(R)$ obeying (\ref{eq:vis_dissip}) equation (\ref{eq:1D}) 
can be rewritten as eq. (\ref{eq:irr_flux_mod}) with 
\ba
g(\theta)=\frac{3\cos\theta}{4\pi^2}
\int\limits_{x_{in}}^\infty \frac{f(xR_\star)q(x,\theta)dx}{x^2 D^2}
\label{eq:g}
\ea
Integral in (\ref{eq:g}) is dominated by
$x\sim x_{in}$. Near equator, where $\theta\to \pi/2$ one can expand 
integrand in (\ref{eq:1D}) in terms of $\pi/2-\theta\ll 1$ and 
$x-1\ll 1$ which results in $F_d(\pi/2)=F_d(R_\star)/2$. 
Polar regions of the star ($\theta\to 0$) are illuminated only by
distant parts of the disk, $R\gtrsim R_\star/\sin\theta\gg 
R_\star$, so that $x\gg 1$ (while $x\sin\theta\sim 1$)
in equation (\ref{eq:1D}). Also, far from the star one can safely 
use equation (\ref{eq:vis_dissip}) with $f=1$ to finally arrive at the 
equation (\ref{eq:as}) with 
\ba
I_1=\frac{3}{4\pi^2}\int\limits_1^\infty\frac{dt}{t^6}
\left(\sqrt{t^2-1}-2\arctan\frac{\sqrt{t^2-1}}{1+t}\right)=\frac{1}{50\pi^2}
\label{eq:I}
\ea
This result is independent of the structure of the boundary layer near
the stellar surface since the polar regions of the star do not have direct
sight lines to the boundary layer. This is evidenced by the convergence 
at $\theta \to 0$ of the two curves in Figure \ref{fig:irr_flux}
calculated assuming $f(R)=1$ and $f(R)=1-(R_\star/R)^2$.


\section{Validity of 1D approximation.}
\label{ap:1D_validity}


To determine the validity limits of the 1D solution for the
structure of the radiative zone found in \S \ref{sect:1D} we evaluate 
the magnitude of the corrections arising when the latitudinal 
radiative transfer is accounted for. Considering 1D solution 
(\ref{eq:PofT}) as a zeroth-order approximation we plug it into
the full equation (\ref{eq:rad_tran}) and carefully expand all 
$\theta$-derivatives, remembering that $P$ is almost independent 
of $\theta$ (latitudinal pressure gradients are small). Integrating
the resultant expression once over $r$ we  again arrive 
at the equation (\ref{eq:flux}) but with 
$F_{in}\to F_{in}+\delta F_{in}$ in the left-hand side, where
\ba
\delta F_{in}=\left(\frac{k_B T_{ph}^{4-\beta}}
{\mu L_\theta^2}\right)^2 g^{-1}
\int\limits_P^{P_{cb}}\frac{T^{\beta-2}}{P^{\alpha+2}}
Z(P)dP.
\label{eq:deltaF}
\ea
Here $Z(P)\sim 1$ is a weak function of pressure (varying by at
most a factor $\sim 1$) and $L_\theta$ 
is a characteristic scale of latitudinal variation of $T_{ph}$,
$L_\theta = R_\star(\partial \ln T_{ph}/\partial \theta)^{-1}$.
Our 1D approximation is justified if the correction 
to the 1D result $\delta F_{in}$ is small compared to $F_{in}$
given by equation (\ref{eq:F_in}). 

Integral in (\ref{eq:deltaF}) attains its 
highest value at $P\sim P_{ph}$ (latitudinal radiation transfer 
is easiest in the upper, low density layers of the star just below
the photosphere) and one can easily find using equations 
(\ref{eq:P_ph}), (\ref{eq:nab_ph}), and (\ref{eq:F_in}) that 
\ba
\frac{\delta F_{in}}{F_{in}}\sim \left(\frac{H_{ph}}
{L_\theta}\right)^2\nabla_{ph}^{-1},
\label{eqflux_ratio}
\ea
where $H_{ph}$ is a photospheric scale height. This result makes
it clear that the 1D solution for the structure of the radiative 
zone should be reasonable as long as the condition (\ref{eq:validity}) 
is fulfilled.


\end{document}